\documentclass[sigconf]{aamas} 

\usepackage{balance} 
\usepackage{amsmath}
\usepackage[inline]{enumitem}
\usepackage{multirow,array}
\usepackage{amsthm}
\usepackage{comment}
\usepackage{bm}
\usepackage{caption}
\usepackage{subcaption}

\setcopyright{ifaamas}
\acmConference[AAMAS '24]{Proc.\@ of the 23rd International Conference
on Autonomous Agents and Multiagent Systems (AAMAS 2024)}{May 6 -- 10, 2024}
{Auckland, New Zealand}{N.~Alechina, V.~Dignum, M.~Dastani, J.S.~Sichman (eds.)}
\copyrightyear{2024}
\acmYear{2024}
\acmDOI{}
\acmPrice{}
\acmISBN{}

\acmSubmissionID{589}

\title[
Reputation and Emergent Cooperation in Social Dilemmas
]{
Emergent Cooperation under Uncertain Incentive Alignment
}

\author{Nicole Orzan}
\affiliation{
  \institution{University of Groningen}
  \city{Groningen}
  \country{The Netherlands}}
\email{n.orzan@rug.nl}

\author{Erman Acar}
\affiliation{
  \institution{ILLC \& IvI, University of Amsterdam}
  \city{Amsterdam}
  \country{The Netherlands}}
\email{erman.acar@uva.nl}

\author{Davide Grossi}
\affiliation{
  \institution{University of Groningen}
  \city{Groningen}
  \country{The Netherlands}}
\email{n.orzan@rug.nl}

\author{Roxana R\u{a}dulescu}
\affiliation{
  \institution{Vrije Universiteit Brussels, Belgium}
 \country{Utrecht University, The Netherlands}}
\email{roxana.radulescu@vub.be}

\begin{abstract}
Understanding the emergence of cooperation in systems of computational agents is crucial for the development of effective cooperative AI. Interaction among individuals in real-world settings are often sparse and occur within a broad spectrum of incentives, which often are only partially known. In this work, we explore how cooperation can arise among reinforcement learning agents in scenarios characterised by infrequent encounters, and where agents face uncertainty about the alignment of their incentives with those of others. To do so, we train the agents under a wide spectrum of environments ranging from fully competitive, to fully cooperative, to mixed-motives. Under this type of uncertainty we study the effects of mechanisms, such as reputation and intrinsic rewards, that have been proposed in the literature to foster cooperation in mixed-motives environments. Our findings show that uncertainty substantially lowers the agents' ability to engage in cooperative behaviour, when that would be the best course of action. In this scenario, the use of effective reputation mechanisms and intrinsic rewards boosts the agents' capability to act nearly-optimally in cooperative environments, while greatly enhancing cooperation in mixed-motive environments as well.
\end{abstract}


\newcommand{\BibTeX}{\rm B\kern-.05em{\sc i\kern-.025em b}\kern-.08em\TeX}

\begin{document}


\pagestyle{fancy}
\fancyhead{}


\maketitle 


\section{Introduction}

The question of how cooperation can emerge in human and animal societies \cite{dong2016dynamics, rubenstein2010cooperation}, as well as in systems of artificial agents has been a focus of research for decades \cite{axelrod1981emergence, nowak2004emergence, dafoe2020, challet1997emergence}.
Understanding the mechanisms behind cooperation emergence is fundamental for the development of artificial agents capable of effectively collaborating with both other agents and humans. 
The most significant challenges in the problem of cooperation emergence are encountered in scenarios where agents' incentives exhibit some degree of misalignment, commonly known as mixed-motive environments \cite{schelling1958strategy}. One instance of such environments are social dilemmas, among which the public goods game \cite{andreoni1988free} is the focus of the present paper.

\textit{State of the art.}
Among all the factors that can undermine cooperation emergence, uncertainty is a prominent one. There are two main types of uncertainty at play in social dilemmas: social uncertainty, which refers to uncertainty regarding the behavior of other agents \cite{deutchman2022common, bendor1993uncertainty}; and 
environmental uncertainty, which refers to the variability of resources obtained from the environment \cite{wit1998public, andras2006uncertainty}.



Addressing social uncertainty is the drive behind many of the mechanisms that have been studied, across disciplines, in order to foster cooperation. Direct and indirect reciprocity are among the main social mechanisms that have been demonstrated to facilitate the emergence of cooperation in scenarios with misaligned interests \cite{schmid2021unified}. While direct reciprocity refers to the behaviour of acting cooperatively with individuals who previously acted cooperatively with us, indirect reciprocity refers to cooperating with individuals who behaved cooperatively with others. 
These cooperation-aiding mechanisms have been extensively studied in social psychology \cite{wu2016reputation, emler1990social, shaw2013reputation, PhysRevE.78.026117}, philosophy \cite{origgi2012social, origgi2019reputation}, computational biology \cite{ohtsuki2015reputation} and evolutionary game theory \cite{xia2023reputation, nowak1998dynamics, ohtsuki2004should, ohtsuki2006leading}.  
The mechanism of indirect reciprocity is particularly relevant in systems where encounters among individuals are rather sparse, and agents have to rely on external information to judge others. One of the ways in which indirect reciprocity can be instantiated is by means of reputation mechanisms. Reputation is an indirect measure of the trustworthiness of an individual, usually rooted in the past conduct the individual exhibited with others \cite{origgi2019reputation, origgi2012social}. Research has shown how reputation can lead to faster emergence of cooperation in social dilemmas and, more generally, mixed-motive games, especially if assisted by other cooperation-driving mechanisms, such as partner selection and trust-building \cite{anastassacos2021cooperation, anastassacos2020partner, gyawali2021, bidgoly2023robustness}. 

To the best of our knowledge, however, it is still unclear to what extent such mechanisms remain robust even under conditions of environmental uncertainty, which has been shown to act as a detrimental factor in the provision of public goods in human groups \cite{wit1998public, gangadharan2009experimental, biel1995}. In particular, we are concerned with situations in which computational agents may be uncertain about how (mis-)aligned their interests actually are in the interaction at play: \emph{To what extent cooperation-aiding mechanisms remain effective under environmental uncertainty?} This is the core research question of the paper.

\vspace{.2cm}
\emph{Contributions.} In this work, we focus on the impact of environmental uncertainty in scenarios where agents' incentives are (mis-)\newline aligned to various extents. We investigate the effect of 
different cooperation-aiding mechanisms (reputation, steering agents and intrinsic rewards) in fostering cooperation among reinforcement learning (RL) even under this type of uncertainty.

This paper presents the three following contributions.
{\em First}, we present the outcomes related to the impact of environmental uncertainty on independent reinforcement learning agents trained on the Extended Public Goods Game (EPGG) \cite{orzan2023}. This game enables training across environments with a variety of incentive alignments, while introducing uncertainty on those. We find that uncertainty diminishes agents' cooperation in cooperative and mixed-motive settings while keeping the competitive behaviour in the competitive ones.
{\em Second}, we present an analysis of the independent role of reputation with effective social norms, steering agents and intrinsic rewards in shaping the behaviour of independent Deep Q-network (DQN) agents \cite{mnih2013playing}, trained on the EPGG with uncertainty. We find that the presence of an effective social norm enables agents to leverage reputation, promoting cooperative behaviour in mixed-motive and cooperative environments. This result is also achieved by the integration of intrinsic rewards based on the idea of self-play.
{\em Third}, we analyse the entangled effect of reputation, intrinsic reward and uncertainty on the behaviour learnt by DQN agents. We show that this combination of social mechanisms brings learning agents closer to the optimal behavior in the competitive environments, particularly when steering agents are employed. Moreover, the inclusion of intrinsic rewards boosts cooperation in the cooperative environments when steering agents are not employed.

\section{Related Work}
\label{sec:relatedwork}


\paragraph{Social Dilemmas} \label{subs:socialdilemmas}
Social dilemmas are group settings 
where the conflict between the individual and collective interest can easily lead to cooperation failure, ultimately giving rise to suboptimal outcomes for the group \cite{kollock1998, dawes1980social}. In recent times, these scenarios are being explored using MARL \cite{gronauer2022multi} to gain insight into the conditions under which cooperative behaviour might emerge.
In \cite{radke2022importance} a model that allows agents to optimise multiple objectives in social dilemmas with team settings is developed. 
In \cite{leibo2017}, with the aim of defining more complex environments than matrix games, temporally extended social dilemmas are introduced. In \cite{jaques2019social} authors work on improving communication in sequential social dilemmas setting via intrinsic motivation, by rewarding agents that have high social influence over other agents. In \cite{eccles2019learning} reciprocation is employed to elicit cooperation in otherwise selfish individuals, giving rise to pro-sociality. While the aforementioned studies focus on analysing games with a mixed-motive structure, in this work we aim at investigating the impact of concurrently training over a set of environments with different levels of incentive alignments, using the public goods game as a source of inspiration. 

\paragraph{Reputation and Social Norms}
\label{subs:reputation}
Reputation is a central mechanism to develop cooperation in multi-agent systems. The concept of reputation is strongly tied to the development of trust among agents \cite{mui2002computational, ostrom2003toward, birk2000boosting}, and therefore to partner selection \cite{capraro2016partner, du2011partner, anastassacos2020partner}. The influence of the partner selection mechanism on the emergence of cooperation among independent RL agents is investigated in \cite{anastassacos2020partner}: agents determine their partner selection choice by observing the actions other agents took in previous interactions. Other works showed how decentralised trust can improve the consensus rate among learning agents in the presence of unreliable agents \cite{fung2022trust}. Combined effect of the mechanisms such as reputation, partner selection, and (direct and third-party) punishment on learning agents has also been studied, showing how the highest rates of cooperation are reached by these  mechanisms simultaneously \cite{dasgupta2023investigating}. In \cite{mckee2023multiagent}, authors investigate intrinsic motivation for reputation in agents trained on spatially and temporally complex tasks, by directly modifying the reward signal. They observe an increase in group coordination in the game when agents are motivated to achieve a good reputation.

Studies in computational and evolutionary biology \cite{ehrlich2005evolution, chalub2006evolution}, evolutionary game theory \cite{voss2001game, matsui1996cultural, ohtsuki2004should, ohtsuki2006leading, nowak1998dynamics}, and artificial intelligence investigated the evolution of social norms, their impact on the group dynamics, and their role in sustaining cooperation. For example, \cite{santos2016social} examines the changes in cooperativity within a group when different social norms are applied. The findings show that the most effective norms are the ones expressing that agents should cooperate with good individuals and defect with bad ones. This is widely observed result (see \cite{ohtsuki2006leading, pacheco2006stern, sen2007}, among others). In \cite{vinitsky2023learning} author investigate the effect of public sanctioning on the emergence of social norms, to solve social dilemmas. In \cite{anastassacos2021cooperation}, the authors study the interaction of reputation mechanisms and social norms in a reinforcement learning framework, and show that the reputation mechanism alone is not sufficient to guide tabular RL agents towards a cooperative equilibrium. However, this approach becomes effective when combined with intrinsic rewards, or when (non-learning) agents that already adhere to the norm are introduced in the system.

In the real-world, interactions are characterised by environments that present different levels of incentive alignment. Furthermore, oftentimes it is not straightforward to determine to which extent the incentives between agents are actually aligned: agents may not be aware of whether they are playing in a cooperative, competitive, or mixed scenario.
All of the aforementioned studies focus on applying social mechanisms on agents that are acting in mixed-motive environments in which the achievement of cooperative behaviour by learning is not guaranteed. However, none of these studies focuses on learning across a range of environments with different levels of incentive alignment, which is the main distinctive character of our work. We are investigating, in particular, the effect of social mechanisms --- building upon the framework of \cite{anastassacos2021cooperation} --- in environments that exhibit different levels of incentive alignments and in presence of uncertainty. We should also note that, in contrast \cite{anastassacos2021cooperation}, we go beyond tabular learning approach, to accommodate the continuous nature of EPGG.

\section{Preliminaries}
\label{sec:preiminaries}

In this section, we introduce (1) the Extended Public Goods Game, which is the main game-theoretic setting of our work, and (2) the key social mechanisms we used to study emergence of cooperation: reputation, social norms and intrinsic rewards.

\subsection{The Extended Public Goods Game}
\label{subs:epgg}

The Public Goods Game (PGG) is a collective decision game extensively studied in economics, which describes situations where cooperation by rational agents is Pareto optimal, but because of the profitability of free-riding, rational agents converge to defection, which is the Nash equilibrium \cite{andreoni1988free}. We refer to this kind of games as mixed-motive, since the incentives of the agents are partially misaligned. 
\begin{figure*}
\bigskip
\centering
\begin{subfigure}[t]{0.49\columnwidth}
\begin{tabular}{cc|c|c|}
  $f=0.5$ & \multicolumn{1}{c}{} & \multicolumn{2}{c}{Player $X$}\\
  & \multicolumn{1}{c}{} & \multicolumn{1}{c}{$C$}  & \multicolumn{1}{c}{$D$} \\\cline{3-4}
  \multirow{2}*{Player $Y$}  & $C$ & $2,2$ & $1,5$ \\\cline{3-4}
  & $D$ & $5,1$ & $4,4$ \\\cline{3-4}
\end{tabular}
\end{subfigure}
\begin{subfigure}[t]{0.49\columnwidth}
\begin{tabular}{cc|c|c|}
  $f=1.0$ & \multicolumn{1}{c}{} & \multicolumn{2}{c}{Player $X$}\\
  & \multicolumn{1}{c}{} & \multicolumn{1}{c}{$C$}  & \multicolumn{1}{c}{$D$} \\\cline{3-4}
  \multirow{2}*{Player $Y$}  & $C$ & $4,4$ & $2,6$ \\\cline{3-4}
  & $D$ & $6,2$ & $4,4$ \\\cline{3-4}
\end{tabular}
\end{subfigure}
\begin{subfigure}[t]{0.49\columnwidth}
\begin{tabular}{cc|c|c|}
  $f=1.5$ & \multicolumn{1}{c}{} & \multicolumn{2}{c}{Player $X$}\\
  & \multicolumn{1}{c}{} & \multicolumn{1}{c}{$C$}  & \multicolumn{1}{c}{$D$} \\\cline{3-4}
  \multirow{2}*{Player $Y$}  & $C$ & $6,6$ & $3,7$ \\\cline{3-4}
  & $D$ & $7,3$ & $4,4$ \\\cline{3-4}
\end{tabular}
\end{subfigure}
\begin{subfigure}[t]{0.49\columnwidth}
\begin{tabular}{cc|c|c|}
  $f=3.5$ & \multicolumn{1}{c}{} & \multicolumn{2}{c}{Player $X$}\\
  & \multicolumn{1}{c}{} & \multicolumn{1}{c}{$C$}  & \multicolumn{1}{c}{$D$} \\\cline{3-4}
  \multirow{2}*{$\hspace{3mm}$ Player $Y$}  & $C$ & $14,14$ & $7,11$ \\\cline{3-4}
  & $D$ & $11,7$ & $4,4$ \\\cline{3-4}
\end{tabular}
\end{subfigure}
\caption{Normal form games instantiating the EPGG for two players (X and Y) with 4 coins each, for four possible values of the multiplication factor: $f=0.5$ (competitive game), $f=1.0$ (boundary game with both $CC$ and $DD$ optimal), $f=1.5$ (mixed-motive game) and $f=3.5$ (cooperative game).} 
\label{fig:games}
\bigskip
\end{figure*}
We model Public Goods Games as a tuple $\langle N, \bm{c}, A, f, \bm{u} \rangle$, where $N$ symbolises the set of players, and $|N| = n \in \mathbb{N}$ is the number of players. Every player $i$ is endowed with some amount of wealth (or coins) $c_i \in \mathbb{R}^{\geq 0}$, and $\bm{c} = (c_1, \ldots, c_n)$ denotes the tuple containing all agents' coins. In this work we will be assuming all endowments to be equal, so for all $i,j \in N$, $c_i = c_j$. Each agent can decide weather to cooperate (investing in the public good) or defect (keeping the endowment); therefore, the set $A$ of allowed actions consists of cooperate $(C)$ and defect $(D)$ i.e., $A = \{C, D\}$. The vector $\bm{a} = (a_1, \dots , a_n) \in A$ represents the action profile of the agents. The number $f \in (1, n)$ is called the multiplication factor, and specifies the quantity by which the total good collected by the agents' investment is multiplied. This resulting quantity generates the public good,  which is then evenly distributed among all agents. We define by $\bm{u}$ the vector of utilities received by the agents, and  $u_i: A^{n} \times (1,n) 
\times ( \mathbb{R}^{\geq 0} )^n 
\rightarrow \mathbb{R}$ is the utility function for agent $i$, specified as follows:

\begin{equation}
\label{eq:utility}
    u_i(\bm{a}, f, \bm{c}) = \frac{1}{n} \sum_{j=1}^{n} c_j I(a_j) \cdot f + c_i (1 - I(a_i)),
\end{equation}
where $a_j$ is the $j-$th entry of the action profile ${\bm a}$ and $I(a_j)$ is the indicator function, equal to 1 if the action of the agent $j$ is cooperative, and 0 otherwise, and $c_j$ denotes the $j-$th entry of $\bm{c}$. 
Because we assumed $1 < f < n$, it is easy to check that, in PGG, the profile with all agents defecting is a dominant strategy equilibrium, while the Pareto optimal profile is the one in which all agents cooperate. This captures the social-dilemma nature of PGG. In particular, for $n = 2$ the PGG defines a class of Prisoner Dilemma games. 

Following \cite{orzan2023}, in this paper we generalise the definition of PGG by letting $f$ range over $(0, R_{+})$ instead of $(1, n)$, where $R_{+} > n$ is an arbitrary value. This defines the class of Extended Public Good Games (EPGG). 
EPGG can model not only mixed-motives scenarios (i.e., when $1 < f \leq n$) but also (1) fully competitive scenarios (i.e., when $0 \leq f < 1$) in which the defection profile is a Pareto optimal dominant strategy equilibrium (thus a Nash equilibrium) and incentives are therefore fully misaligned; and (2) fully cooperative scenarios (i.e., when $n \leq f \leq R_{+}$) in which the cooperation profile is a Pareto optimal dominant strategy equilibrium (thus an optimal Nash equilibrium) and incentives are therefore fully aligned. Note that in the boundary case $f=1$ both CC and DD are Pareto optimal but only DD is an equilibrium. It is important to mention already that our agents will be trained in the EPGG under uncertainty about the value of $f$. The uncertainty of agent $i$ consists of an observed value of the multiplication factor $f^{obs}_i \in (0, R_{+})$ which, however, may differ from the true one (i.e., $f$). As detailed in Section \ref{sec:uncertainty}, we model the uncertainty on the observation as Gaussian noise over the true value of $f$ (while ensuring that any observed value $f^{obs}_i$ less than 0 is truncated to 0).

In our work we consider interactions among two agents at a time. Figure \ref{fig:games} shows the normal form of the matrix games instantiating the EPGG for $2$ players, with $N = \{X,Y\}$ endowed with $4$ coins each, for the following set of multiplication factors: $f \in \{0.5, 1.0, 1.5, 3.5\}$. These games model fully competitive ($f \in \{0,1\}$), fully cooperative ($f \in \{3.5\}$) and mixed-motives ($f = 1.5$) interactions.

\subsection{Reputation}
\label{subs:reputationmechanism}

Following~\cite{anastassacos2021cooperation, ohtsuki2004should}, we model reputation for an agent $i$ as a binary variable, $r_i\in\{0, 1\}$ in which 1 (resp. 0) will denote \emph{good-} (resp. \emph{bad-}) \emph{reputation}. How these values are assigned depends on {\em social norms}, mechanisms that drive reputation update in the group.
Social norms are basically functions i.e., $g: A \times \{0,1\} \to \{0,1\}$ that update the reputation of each playing agent after every interaction. That is, the new value of the reputation $r_i$ is computed based on the last action $a_i$ took by the agent, and the current reputation $r_j$ of the opponent i.e., $r_i = g(a_i, r_j)$.
Given two possible values for both the reputation and the available actions, there are $2^{2^2}$ possible social norms that could be defined \cite{ohtsuki2004should}. It has been shown that some social norms are more apt to push and maintain cooperation than others \cite{ohtsuki2006leading}. One of the most effective social norms for sustaining cooperation in mixed-motive scenarios is to assign a good reputation to those who cooperate with agents having good reputations and defect against those with bad reputations, regardless of their actions. Formally, 
\begin{align} \label{eq:norm}
    g(a_i,r_j) = 
    \begin{cases}
        1 & \mbox{if} \ a_i = C \ \mbox{and} \ r_j = 1, \\
        1 & \mbox{if} \ a_i = D \ \mbox{and} \ r_j = 0, \\
        0 & \mbox{otherwise}
    \end{cases}
\end{align}
i.e., a reputation of 1 is assigned to an agent that cooperates with an opponent that has reputation $1$, or to an agent that defects against an opponent with reputation $0$, and assigns 0 otherwise. 


The social norm defined in Equation \eqref{eq:norm} is the norm we deploy in our experiments. However, application of social norms to the EPGG requires one further adaptation with respect to existing approaches. Because of the uncertainty about the multiplicative factor, we need to consider the fact that agents should not be expected by the norm to cooperate when operating in fully competitive environments. To account for this, we make the social norms dependent on the multiplication factor of the game that is currently being played. So, social norms for the EPGG are functions 
$G: [0,R_{+}] \times A \times \{0,1\}\times \{0,1\} \to \{0,1\}$. 
In this setting we will be working with one specific adaptation of Equation \eqref{eq:norm}:
\begin{align} \label{eq:norm2}
    G(f,a_i,r_i,r_j) = 
    \begin{cases}
        g(a_i,r_j) & \mbox{if } f \geq 1, \\
        r_i & \mbox{otherwise}
    \end{cases}
\end{align}
That is, if the multiplication factor currently played is bigger or equal than $1$ --- the boundary between a competitive ($f < 1$) and a mixed motive ($f > 1$) interaction --- we let the social norm modify the reputation of agent $i$ as described earlier. However if the multiplication factor is less than $1$, reflecting a competitive setting, the reputation of the agent stays fixed at its current value $r_i$. 
Such design is justified as, in competitive games, the cooperative action does not represent a Pareto optimal outcome nor a Nash equilibrium. Consequently, neither the group nor the individual would benefit from it. However, in the boundary case $f=1$ --- where both full cooperation and full defection are Pareto optimal (see Figure \ref{fig:games}) --- the norm demands an out-of-equilibrium response attributing positive reputation to cooperation rather than defection. 
With the introduction of reputation, the policy of an RL agent $i$ playing on the EPGG against an opponent $j$ depends both on the observed value of the multiplication factor $f^{obs}_i$, which can be affected by uncertainty, and the reputation of the opponent $r_j$. Thus, the policy of agent $i$ is a function $\pi_i: [0,R_{+}] \times \{0,1\} \rightarrow [0,1]$. The action taken by agent $i$ is then sampled from the distribution: 
$a_i \sim \pi_i(\cdot|f^{obs}_i, r_j)$. 
Besides the social norms driving the reputation update, reputation systems may also benefit from bootstrapping effects provided by agents with fixed behaviors. In \cite{anastassacos2021cooperation} authors show that relying solely on the reputation mechanism is insufficient to shift the equilibrium of the Prisoner's Dilemma towards cooperation. 
However, they also noted that introducing agents that do not respond to the incentives of the game but rather exhibit fixed behaviour, strictly adhering to specific social norms, may significantly enhance cooperation among agents. They refer to such agents as \emph{steering agents}. In our study we leverage this approach, experimenting with different proportions of steering agents in the game, to evaluate the impact of those on the learning agents. Following \cite{anastassacos2021cooperation} we implement a steering agent $i$ as an agent that follows the social norm defined by Equation \eqref{eq:norm2}. Noting that our objective is investigating the agent's behaviour in the presence of observation uncertainty, we emphasise that steering agents receive uncertain observations as well. 
A steering agent $i$ playing against an opponent $j$ with reputation $r_j$, receiving observation $f^{obs}_i$ on the value of the multiplication factor, employs the fixed policy $\pi_s$:
\begin{align} \label{eq:steering_policy}
    \pi_s(f^{obs}_i,r_j) = 
    \begin{cases}
        1 & \mbox{if } f^{obs}_i \geq 1 \mbox{ and } r_j = 1\\
        0 & \mbox{otherwise}
    \end{cases}
\end{align}

\subsection{Intrinsic Reward}
\label{subs:intrinsicreward}

The idea of intrinsic reward comes from the psychological concept of intrinsic motivation \cite{fishbach2022structure}, which characterises the drive of doing something for its inherent enjoyability rather than for any external or utilitarian gain \cite{chentanez2004intrinsically}. This defines a reward signal that does not come from an external source (i.e., the environment), but from the agent itself \cite{barto2013intrinsic}. 
Following the model provided by \cite{anastassacos2021cooperation}, we ground intrinsic rewards in the idea of self-play, namely, those are the utility that agent $i$ \emph{imagines} it would obtain when playing against itself. 
This choice is motivated by the goal of prioritising learning policies that are effective against agents with a similar behaviour. Intrinsic reward is defined as the utility that agent $i$, with reputation $r_i$, would receive when playing against an agent with its same reputation in an environment where the true multiplication factor is the observed one: $u_i(\bm{a}^{\prime}, f^{obs}_i, \bm{c})$, where $\bm{a}^{\prime} = (a^{\prime}_i, a^{\prime}_i)$ and $a_i \sim \pi_i(\cdot|f^{obs}_i, r_i)$.

When the intrinsic reward formulation is employed, the reward function $R_i: A^{n} \times \mathbb{R}^{\geq 0} \times \mathbb{R}^n \rightarrow \mathbb{R}$ of agent $i$ playing against agent $j$ on a game with multiplication factor $f$ and receiving observation $f^{obs}_i$, is a convex combination of the two utilities:
\begin{equation}
\label{eq:intrinsic}
    R_i = \beta \; u_i(\bm{a}, f, \bm{c}) + (1-\beta) \; u_i(\bm{a}^{\prime}, f^{obs}_i, \bm{c}),
\end{equation}

\noindent where $\bm{a} = (a_i, a_j)$ and $a_i \sim \pi_i(\cdot|f^{obs}_i, r_j)$, and $u_i(\bm{a}, f, \bm{c})$ is the utility received from the EPGG. $\beta \in [0,1]$ is a parameter that determines the weight of the game utility $u_i(\bm{a}, f, \bm{c})$ against the intrinsic reward $u_i(\bm{a}^{\prime}, f^{obs}_i, \bm{c})$ i.e.,  when $\beta=1$ (resp. $\beta=0$) the obtained reward is the game utility (resp. intrinsic reward). 

\section{Methods}
\label{sec:methods}

In this section we present the experimental framework we employ to investigate the effect of reputation, steering agents and intrinsic rewards on the behaviour of independent reinforcement learning agents trained on the EPGG. To assess the effect of these elements on the system, we conduct four different kind of experiments, considering both scenarios with and without uncertainty in the agents' observation of the multiplication factor.
{\em First}, we perform a study where the reputation and the intrinsic rewards mechanism are disabled. {\em Second}, we enable the reputation mechanism. In these experiments we additionally investigate the influence of the presence of steering agents to enforce norms within the system. To do so, we observe the results on the average cooperation---that is, the average relative frequency with which cooperative actions are selected by the agents---under different proportions of steering agents. {\em Third}, we perform experiments where only the intrinsic reward mechanism is enabled. And {\em fourth}, we study the interaction between reputation mechanism, steering agents and intrinsic rewards.

As described in Section \ref{subs:epgg}, the EPGG consists of a set of games, each characterised by a multiplication factor $f$, that is a continuous variable in $(0, R_{+})$. In order to compare our results to previous work on reputation mechanisms and MARL \cite{anastassacos2021cooperation}, we opt for tabular Q-learning \cite{watkins1992q} as a baseline. In this case, the set of multiplication factors needs to be discretised. To alleviate this restriction, we employ Deep Q-Networks (DQN) \cite{mnih2013playing}, as the natural function approximation extension of the tabular approach. To the best of our knowledge, ours is the first study of reputation mechanisms for emergent cooperation with DQN agents.
\paragraph{Training process} To conduct the experiments, we employ a pool of $10$ agents whose endowments are all fixed to the same constant value $c = 4$. At every training epoch two players (that will be referred to as active agents) are uniformly sampled from the pool, and a multiplication factor $f$ is sampled uniformly from a predefined set of values. Afterwards, the two selected agents play repeatedly the matrix game defined by $f$ for a fixed number of rounds $M$. 
At each of these rounds every agent $i$ observes the current multiplication factor of the game $f^{obs}_i$, potentially with uncertainty (how precisely this uncertainty is modelled we will deal with in Section \ref{sec:uncertainty}). When no uncertainty is introduced, the observed value of the $f$ coincides with the true value: $f^{obs}_i = f, \; \forall \; i \; \in N$. If the reputation mechanism is not enabled, this observed value is the only input the active agents receive. In this case the agent's policy takes the simplified form $\pi_i : [0, R_{+}] \rightarrow [0,1]$, and $a_i \sim \pi(\cdot|f^{obs}_i)$. 
If the reputation mechanism is enabled, the agents receive as part of the input the reputation $r_j$ of the opponent. The type of the policy in this case is the one defined in Section \ref{subs:reputationmechanism}.

Upon receiving their observations, the active agents take a cooperate or defect action according to their policy. 
After acting, the agents receive a reward from the environment, which, if the intrinsic reward mechanism is not enabled, is equal to the utility function of the EPGG for the current game being played. Otherwise, the reward function is of the form described in Section \ref{subs:intrinsicreward}. 
Subsequently, the reputation of each active agent $i$ is updated depending on the actual multiplication factor $f$, the action taken ($a_i$), the opponent's reputation ($r_j$), 
in accordance with the social norm in 
Equation \eqref{eq:norm2}.

We refer to the set of EPGG games employed for training by describing the set of associated multiplication factors. 
In the non-tabular case (Deep Q-learning \cite{mnih2013playing}), agents are trained on the continuous range of values $F_1 = [0.5, 3.5]$. In the tabular case (Q-learning \cite{watkins1992q}) the set $F_2 = \{0.5, 1.0, 1.5, 3.5\}$ has been employed. This second set has been chosen because it contains games with cooperative ($f=3.5$), competitive ($f=\{0.5, 1.0\}$) and mixed-motive nature ($f=\{1.0, 1.5\}$), while $F_1$ is the continuous extension of $F_2$. 

The exploration strategy employed by the agents is $\epsilon-$greedy, while the evaluation policy selects greedily the action that maximises the Q-function. Agents are trained independently for $10.000$ epochs, and the number of interactions between the two active agents has been fixed to $M=200$. We also set the learning rate to $\alpha=0.01$, discount factor $\gamma=0.99$ and the reputation assignment error $\chi=0.001$ (in accordance with \cite{anastassacos2021cooperation}). When intrinsic rewards are enabled, we fixed $\beta=0.1$. 
When using Q-learning, replicating the setup of \cite{anastassacos2021cooperation}, agents collect in a memory buffer all the $M$ interactions that occur at the current epoch. The Q-learning update is implemented only after all the $M$ interactions are completed, leaving the policy unaltered in this period. The interactions stored in the memory buffer are then discarded after the update, to account both for the possible change in the game played and the non-stationarity of the environment. For training the Q-learning algorithm, we set the current exploration rate $\epsilon$ to $0.01$. 
The architecture used for the DQN algorithm consists of a Multi-Layered Perceptron with one hidden layer of size 4 and a ReLU activation function. For training the DQN algorithm we employ a decaying exploration rate, with a starting value of $\epsilon=0.1$ and an ending value of $\epsilon=0.001$. As for the Q-learning update, the DQN update takes place for the two playing agents at the end of the $M$ interactions. The memory buffer is cleared after every interaction.

\paragraph{Evaluation}
The metric we analyse is the average cooperation of the active agents after every training epoch, i.e, the average relative frequency of cooperative actions. To assess the statistical significance between the results of the different experiments, the Welch’s t-test \cite{welch1947generalization} with a p-value of $0.0001$ has been employed, where the variable of interest is the average cooperation of the active agents over the last 50 training epochs of every run, also averaged over 20 runs. For both the tabular and non-tabular cases, the evaluation has been performed on the set $F_2$.

\section{Learning without Uncertainty}
\label{sec:repinepgg}

In this section, we present the baseline results of the effect of reputation, steering agents, and intrinsic rewards on the behavior of independent RL agents trained on the EPGG without uncertainty on $f$, so with perfect knowledge on the incentives (mis-)alignment.\footnote{Due to space constraints, we present results and figures for the DQN algorithm, while the results for Q-learning are discussed, but the visualisation of the outcomes is available in the supplementary material.
}

\paragraph{Q-learning} 
In the absence of any cooperation-aiding mechanisms, Q-learning agents quickly converge to the rational behaviour described in Section \ref{subs:epgg}: in the cooperative game ($f=\{3.5\}$) agents converge to cooperation, while in the competitive and mixed motive games ($f=\{0.5, 1.0, 1.5\}$), they converge to defection. This result aligns with the baseline findings in \cite{orzan2023}, where two REINFORCE \cite{williams1992simple} agents are trained in a two-players EPGG without uncertainty. When reputation is enabled, cooperation is quickly reached and stably maintained in the cooperative environment but also in the mixed-motive one, without the need of introducing steering agents. At $f=0.5$ defection remains a stable equilibrium. At $f=1.0$ agents do not perfectly converge to defection, even if the rates of cooperation are very low. The introduction of steering agents does not significantly change the outcomes. The employment of the introspective reward mechanism has an impact only at $m=1.5$: here agents do not converge to defection or cooperation, but reach an average cooperation of $0.51 \pm 0.21$. The combination of reputation and intrinsic rewards affects only the threshold environment $f=1.0$ when at least $70\%$ of the population consists of steering agents, shifting the defective equilibria to more cooperative outcomes.

Interestingly, the result for the mixed-motive environment when employing reputation differs from the one obtained by \cite{anastassacos2020partner} on the Prisoner's Dilemma game: here, the authors observe that enabling reputation and effective social norms alone do not suffice to emerge and support cooperation in a system of $10$ agents playing the repeated Prisoner's Dilemma game. However, the difference of the outcome in our experiments is justifiable if we look at the gap between the utilities the agents get in the cooperation-cooperation and defection-defection equilibria: in our setting the gap among these utilities is smaller. We note that our finding is also consistent with the argument that plain tabular reinforcement learning agents can converge to cooperation in mixed-motive games even without the use of reputation and memory \cite{dolgopolov2022reinforcement}.

\begin{figure*}[h]
     \centering
\begin{minipage}[t]{\textwidth}
\centering
    \begin{subfigure}[b]{0.23\textwidth}
         \centering
         \includegraphics[width=\textwidth]{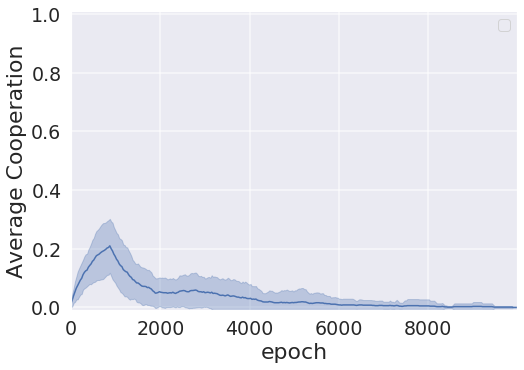}
         \caption{$f=0.5$}
         \label{fig:tr1}
     \end{subfigure}
    \begin{subfigure}[b]{0.23\textwidth}
         \centering
         \includegraphics[width=\textwidth]{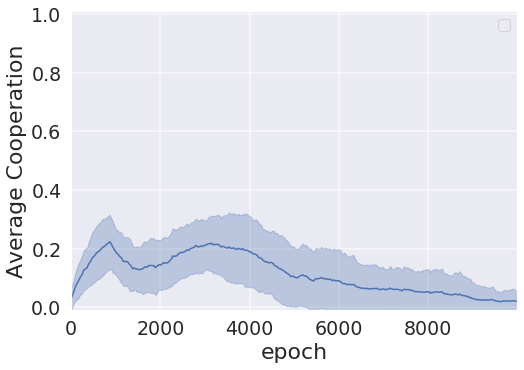}
         \caption{$f=1.0$}
         \label{fig:tr2}
     \end{subfigure}
      \begin{subfigure}[b]{0.23\textwidth}
         \centering
         \includegraphics[width=\textwidth]{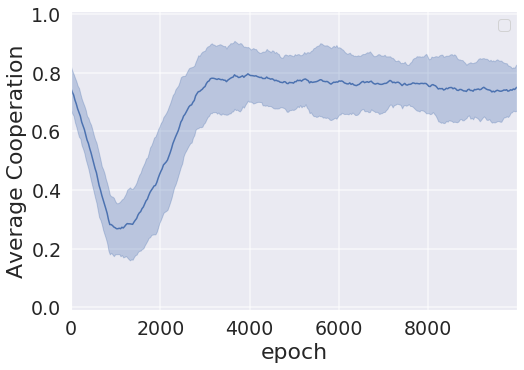}
         \caption{$f=1.5$}
         \label{fig:tr3}
     \end{subfigure}
       \begin{subfigure}[b]{0.23\textwidth}
         \centering
         \includegraphics[width=\textwidth]{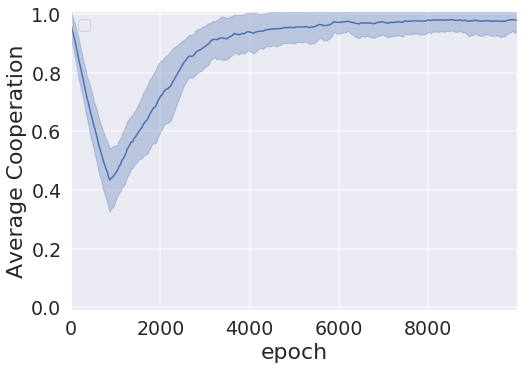}
         \caption{$f=3.5$}
         \label{fig:tr4}
     \end{subfigure}
\end{minipage}
\begin{minipage}[t]{\textwidth}
\centering
     \begin{subfigure}[b]{0.23\textwidth}
         \centering
         \includegraphics[width=\textwidth]{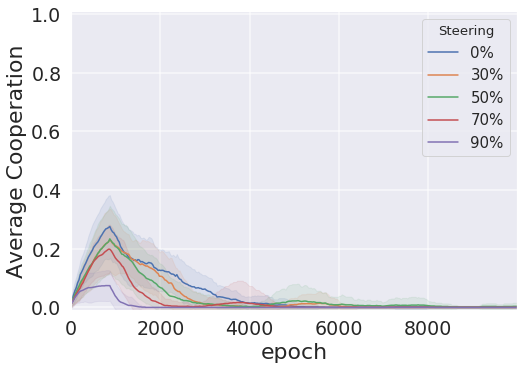}
         \caption{$f=0.5$}
         \label{fig:tr5}
     \end{subfigure}
     \begin{subfigure}[b]{0.23\textwidth}
         \centering
         \includegraphics[width=\textwidth]{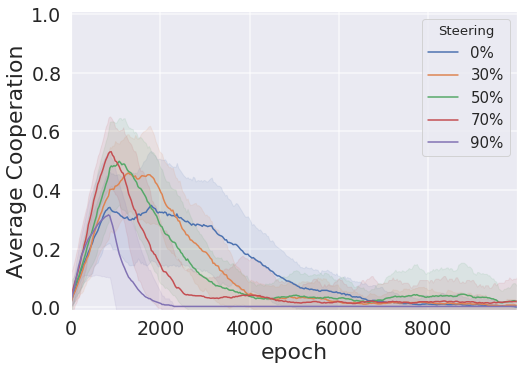}
         \caption{$f=1.0$}
         \label{fig:tr6}
     \end{subfigure}
    \begin{subfigure}[b]{0.23\textwidth}
         \centering
         \includegraphics[width=\textwidth]{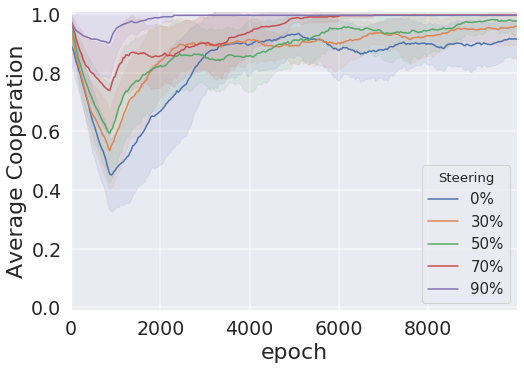}
         \caption{$f=1.5$}
         \label{fig:tr7}
     \end{subfigure}
     \begin{subfigure}[b]{0.23\textwidth}
         \centering
         \includegraphics[width=\textwidth]{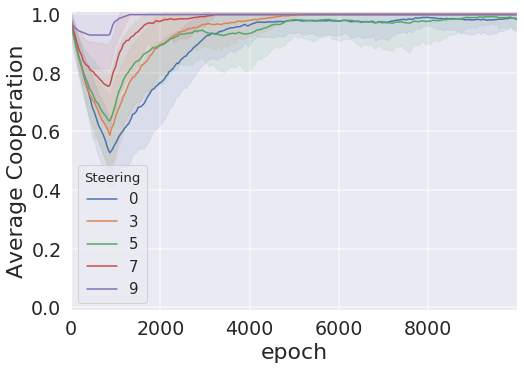}
         \caption{$f=3.5$}
         \label{fig:tr8}
     \end{subfigure}
\end{minipage}
\begin{minipage}[t]{\textwidth}
\centering
     \begin{subfigure}[b]{0.23\textwidth}
         \centering
         \includegraphics[width=\textwidth]{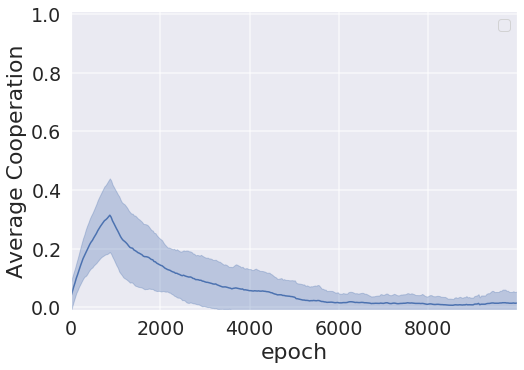}
         \caption{$f=0.5$}
         \label{fig:tr5}
     \end{subfigure}
     \begin{subfigure}[b]{0.23\textwidth}
         \centering
         \includegraphics[width=\textwidth]{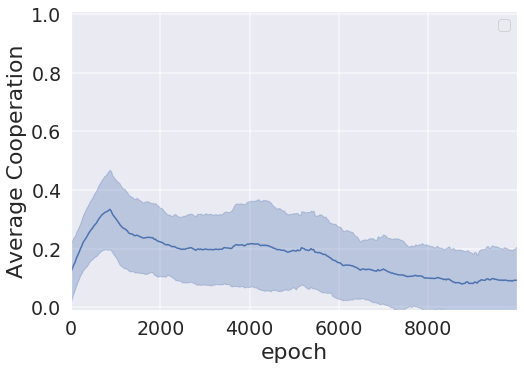}
         \caption{$f=1.0$}
         \label{fig:tr6}
     \end{subfigure}
    \begin{subfigure}[b]{0.23\textwidth}
         \centering
         \includegraphics[width=\textwidth]{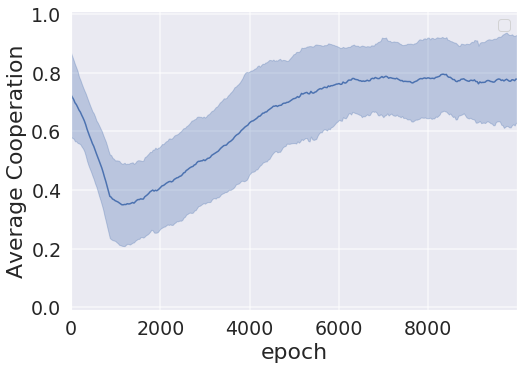}
         \caption{$f=1.5$}
         \label{fig:tr7}
     \end{subfigure}
     \begin{subfigure}[b]{0.23\textwidth}
         \centering
         \includegraphics[width=\textwidth]{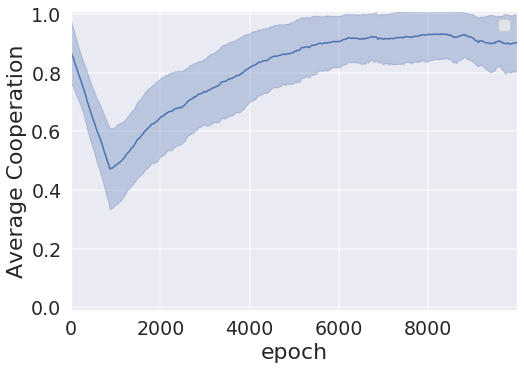}
         \caption{$f=3.5$}
         \label{fig:tr8}
     \end{subfigure}
\end{minipage}
\caption{Average cooperation for DQN agents trained across environments with different multiplication factors. The top row (a - d) shows the results in the absence of cooperation-aiding mechanisms, the middle row (e - h) in the presence of reputation mechanisms and a social norm that aids cooperation, and the bottom row (i - l) in the presence of intrinsic rewards.}
\label{fig:2}
\end{figure*}

\paragraph{Deep-Q Network}
The outcomes of the experiments for the scenario without uncertainty are depicted in Figure \ref{fig:2}. The plots show the average cooperation of the active agents at each evaluation step, averaged over 20 runs. The results with and without reputation mechanism are consistent with those found using Q-learning in the cooperative environment $f=3.5$, the competitive environment $f=0.5$ and the threshold one $f=1.0$. However we can notice that, even in absence of the reputation mechanism, in the mixed-motive environment $f=1.5$ the DQN agents reach an average cooperation value of $0.78 \pm 0.09$, which is significantly far from the competitive Nash equilibria of the game. The source of this outcome might be twofold. Firstly, it may arise from the fact that the DQN algorithm does not ensure convergence to the Nash equilibria. Secondly, the imbalance in interval lengths between the competitive games ($f \in [0.5, 1.0]$) and the cooperative ones ($f \in [2.0, 3.5]$), can be the reason leading to a greater influence of the latter on the mixed-motive environments ($f \in [1.0, 2.0]$).
The middle row of Figure \ref{fig:2} shows that the introduction of reputation, even without steering agents, statistically improves the average cooperation of the mixed-motive environment, leaving the results for the cooperative, the competitive and the threshold environments unchanged\footnote{The statistical significance has been assessed using the Welch’s t-test \cite{welch1947generalization} with a p-value of $0.0001$.}. The incorporation of intrinsic rewards brings about the same outcome as we observe in the absence of any additional mechanisms (further details are provided in the supplementary material). 

\begin{figure*}[h]
     \centering
\begin{minipage}[t]{\textwidth}
\centering
    \begin{subfigure}[b]{0.23\textwidth}
         \centering
         \includegraphics[width=\textwidth]{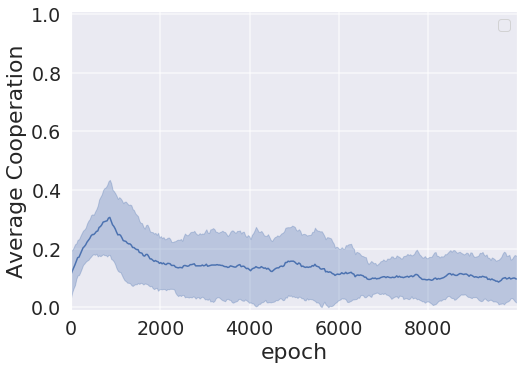}
         \caption{$f=0.5$}
         \label{fig:tr1}
     \end{subfigure}
    \begin{subfigure}[b]{0.23\textwidth}
         \centering
         \includegraphics[width=\textwidth]{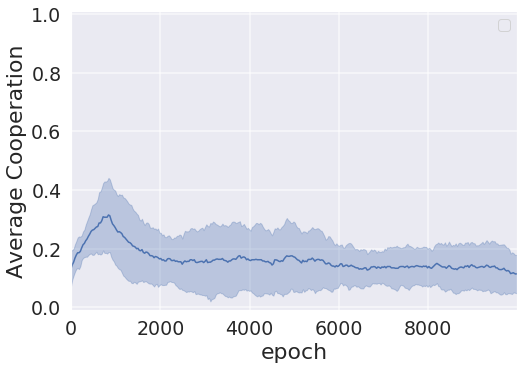}
         \caption{$f=1.0$}
         \label{fig:tr2}
     \end{subfigure}
      \begin{subfigure}[b]{0.23\textwidth}
         \centering
         \includegraphics[width=\textwidth]{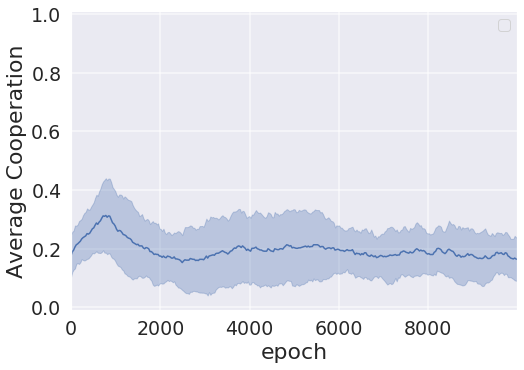}
         \caption{$f=1.5$}
         \label{fig:tr3}
     \end{subfigure}
       \begin{subfigure}[b]{0.23\textwidth}
         \centering
         \includegraphics[width=\textwidth]{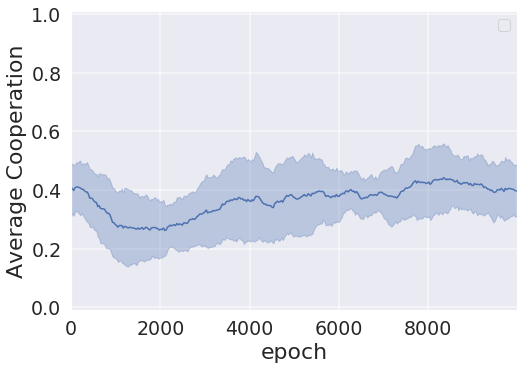}
         \caption{$f=3.5$}
         \label{fig:tr4}
     \end{subfigure}
\end{minipage}
\begin{minipage}[t]{\textwidth}
\centering
     \begin{subfigure}[b]{0.23\textwidth}
         \centering
         \includegraphics[width=\textwidth]{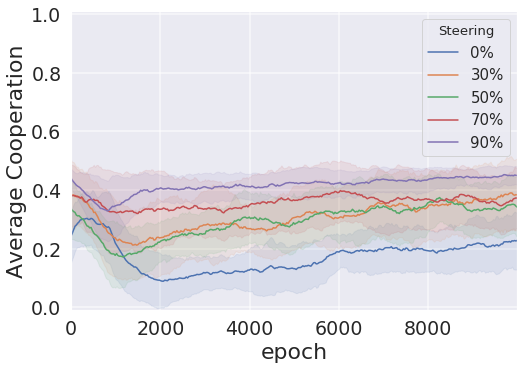}
         \caption{$f=0.5$}
         \label{fig:tr5}
     \end{subfigure}
     \begin{subfigure}[b]{0.23\textwidth}
         \centering
         \includegraphics[width=\textwidth]{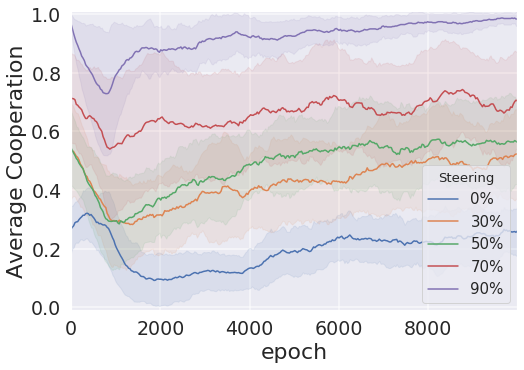}
         \caption{$f=1.0$}
         \label{fig:tr6}
     \end{subfigure}
    \begin{subfigure}[b]{0.23\textwidth}
         \centering
         \includegraphics[width=\textwidth]{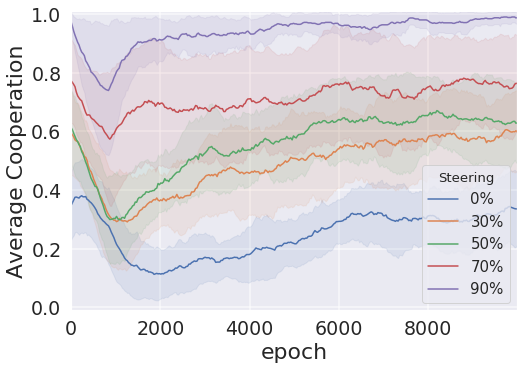}
         \caption{$f=1.5$}
         \label{fig:tr7}
     \end{subfigure}
     \begin{subfigure}[b]{0.23\textwidth}
         \centering
         \includegraphics[width=\textwidth]{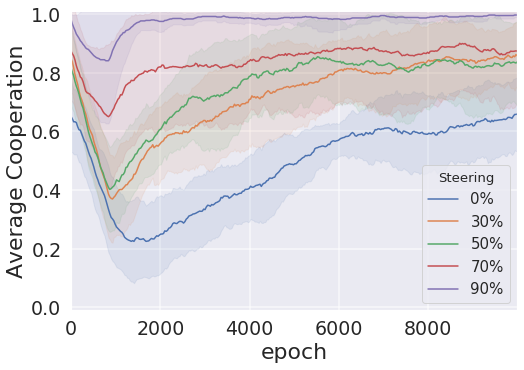}
         \caption{$f=3.5$}
         \label{fig:tr8}
     \end{subfigure}
\end{minipage}
\begin{minipage}[t]{\textwidth}
\centering
    \begin{subfigure}[b]{0.23\textwidth}
         \centering
         \includegraphics[width=\textwidth]{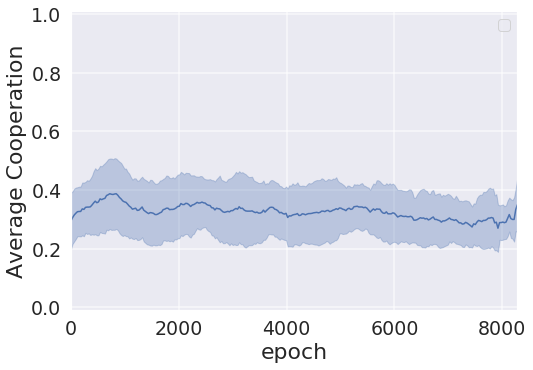}
         \caption{$f=0.5$}
         \label{fig:4tr1}
     \end{subfigure}
    \begin{subfigure}[b]{0.23\textwidth}
         \centering
         \includegraphics[width=\textwidth]{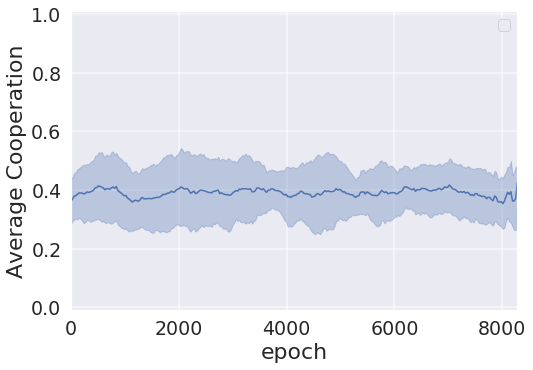}
         \caption{$f=1.0$}
         \label{fig:4tr2}
     \end{subfigure}
      \begin{subfigure}[b]{0.23\textwidth}
         \centering
         \includegraphics[width=\textwidth]{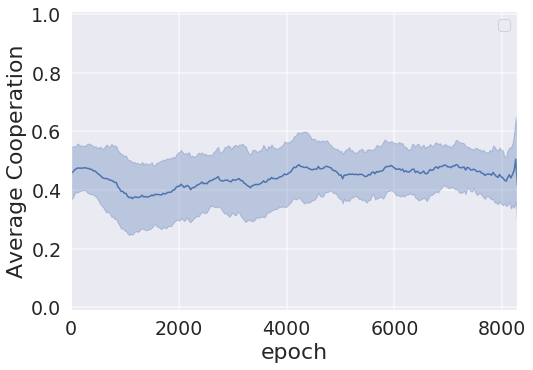}
         \caption{$f=1.5$}
         \label{fig:4tr3}
     \end{subfigure}
       \begin{subfigure}[b]{0.23\textwidth}
         \centering
         \includegraphics[width=\textwidth]{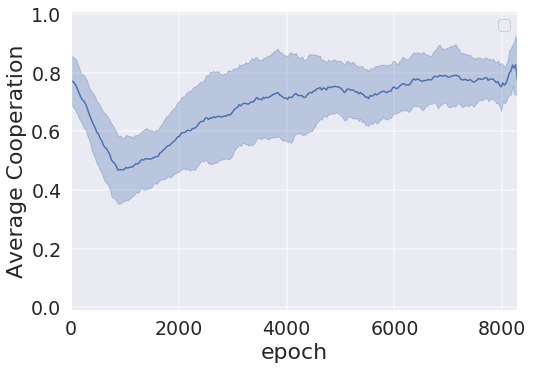}
         \caption{$f=3.5$}
         \label{fig:4tr4}
     \end{subfigure}
\end{minipage}
\begin{minipage}[t]{\textwidth}
\centering
     \begin{subfigure}[b]{0.23\textwidth}
         \centering
         \includegraphics[width=\textwidth]{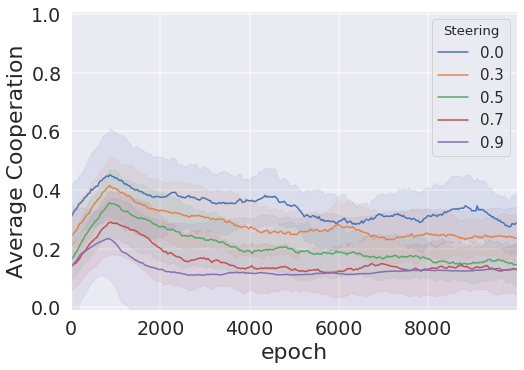}
         \caption{$f=0.5$}
         \label{fig:4tr5}
     \end{subfigure}
     \begin{subfigure}[b]{0.23\textwidth}
         \centering
         \includegraphics[width=\textwidth]{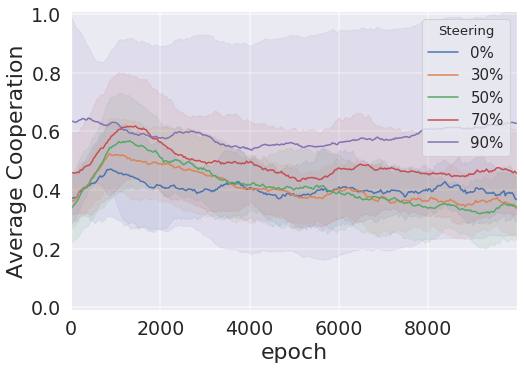}
         \caption{$f=1.0$}
         \label{fig:4tr6}
     \end{subfigure}
    \begin{subfigure}[b]{0.23\textwidth}
         \centering
         \includegraphics[width=\textwidth]{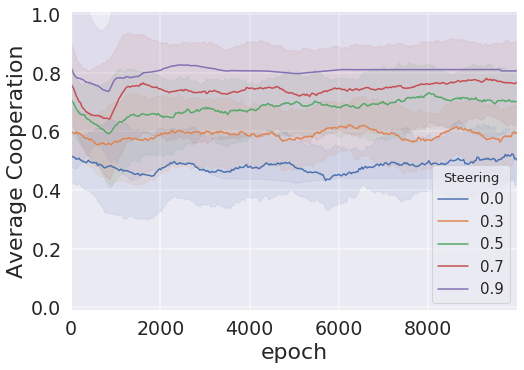}
         \caption{$f=1.5$}
         \label{fig:tr7}
     \end{subfigure}
     \begin{subfigure}[b]{0.23\textwidth}
         \centering
         \includegraphics[width=\textwidth]{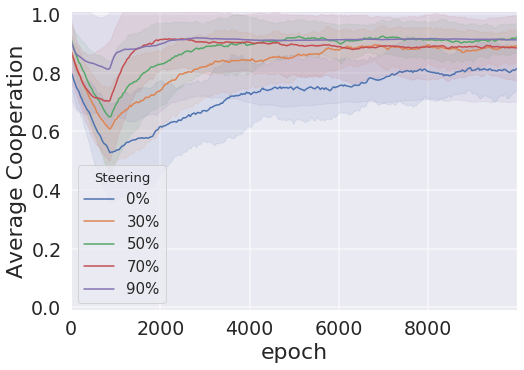}
         \caption{$f=3.5$}
         \label{fig:4tr8}
     \end{subfigure}
\end{minipage}
\caption{Average cooperation for DQN agents trained across environments with different multiplication factors under uncertainty ($\sigma_i = 2 \; \forall \; i \in N$). Results are displayed in four rows: the first without reputation or intrinsic rewards, the second with reputation and a social norm that aids cooperation, the third with the intrinsic rewards formulation, and the fourth with reputation, a social norm that aids cooperation, and the intrinsic rewards formulation.}
\label{fig:3}
\end{figure*}

\section{Learning with Uncertainty}
\label{sec:uncertainty}

In this section, we investigate the role of reputation, steering agents and intrinsic rewards on the behaviour of independent RL agents trained on the EPGG when introducing uncertainty about $f$, so with only noisy information about incentive (mis-)alignment. We model uncertainty over the observation of the multiplication factor $f$ as Gaussian noise: for agent $i$ the observed multiplication factor is sampled from the distribution $f^{obs}_i \sim f + \mathbf{N}(0,\sigma^2_i)$, where $\sigma_i$ is the uncertainty experienced by agent $i$, which is the standard deviation of the distribution. We note that, to ensure the adherence of the observed values to the definition of $f$ inside the EPGG, we clip any sampled value $f^{obs}_i < 0$ to $0$. In these experiments, to allow RL agents to receive continuous observations we utilise the DQN algorithm. The DQN architecture and the employed hyperparameters are identical to the ones previously outlined in Section \ref{sec:methods}. Here we set all agents (both the learning agents and the steering ones) to have the same uncertainty: $\sigma_i = \sigma, \forall \; i \in N$. We fix the value $\sigma=2$. 
In Figure \ref{fig:3} we present the plots for the results of the four experiments outlined in Section \ref{sec:methods} in the presence of uncertainty. These plots depict the average cooperation for each training epoch which, we recall, is computed as an average over 20 runs. Below we provide an overview of the obtained results. For a more in-depth discussion, please refer to Section \ref{sec:depth} of the supplementary material.

\begin{table*}[htbp]
  \centering
  \caption{Averages and standard deviations for the cooperation values computed for each game, in the cases without uncertainty (No U), with uncertainty (U), and with uncertainty and intrinsic rewards (UI), and p-values resulting from the Welch's t-tests.}
    \begin{tabular}{c|c|c|c|c|c}
        \toprule
       Multiplication factor & No Uncertainty & Uncertainty & Uncertainty with I & p-value (No U/U) & p-value  (U/UI)\\
        \midrule
        0.5 & 0.00 $\pm$ 0.02 & 0.09 $\pm$ 0.07 & 0.31 $\pm$ 0.10 & 0.0021 & 2.5337e-8\\
        1.0 & 0.02 $\pm$ 0.04 & 0.12 $\pm$ 0.06 & 0.36 $\pm$ 0.13 & 0.0003 & 1.2248e-9\\
        1.5 & 0.78 $\pm$ 0.09 & 0.16 $\pm$ 0.06 & 0.45 $\pm$ 0.13 & 7.8726e-19 & 3.624e-11\\
        3.5 & 0.98 $\pm$ 0.03 & 0.40 $\pm$ 0.07 & 0.78 $\pm$ 0.12 & 1.7306e-14 & 3.254e-12\\
        \bottomrule
    \end{tabular}
  \label{tab:stat_unc_no_unc}
\end{table*}

\begin{table*}[ht!]
\small
\centering
\caption{Averages and standard deviations for the cooperation in each game, averaged over the last 50 training epochs, for the experiments with uncertainty. The measures are reported separately for two scenarios: the one where only the reputation mechanism is enabled (R) and the one where both reputation mechanisms and intrinsic rewards are employed (RI). Values that exhibit statistical significance are in bold.}
\begin{minipage}[t]{\textwidth}
\centering
\label{tab:rep_intrinsic}
\begin{tabular}{c|cc|cc|cc|cc|cc}
\multirow{2}{*}{}
& \multicolumn{8}{c}{$\hspace{15mm}$ Percentage of Steering Agents}\\
&\multicolumn{2}{c}{\multirow{2}{*}{0\%}}
&\multicolumn{2}{c}{\multirow{2}{*}{30\%}}
&\multicolumn{2}{c}{\multirow{2}{*}{50\%}}
&\multicolumn{2}{c}{\multirow{2}{*}{70\%}}
&\multicolumn{2}{c}{\multirow{2}{*}{90\%}}\\
$f$ &\multicolumn{2}{c}{}&\multicolumn{2}{c}{}\\
\cline{2-11}
& R & RI & R & RI & R & RI & R & RI & R & RI \\
\hline
\multirow{2}*{} 0.5 & 0.22 $\pm$ 0.08 & 0.30 $\pm$ 0.11 & 0.29 $\pm$ 0.09 & 0.23 $\pm$ 0.06 & $\mathbf{0.34 \pm 0.09}$ & $\mathbf{0.15 \pm 0.06}$ & $\mathbf{0.37 \pm 0.08}$ & $\mathbf{0.13 \pm 0.07}$ & $\mathbf{0.45 \pm 0.03}$ & $\mathbf{0.13 \pm 0.20}$ \\
\multirow{2}*{} 1.0 & 0.25 $\pm$ 0.06 & 0.37 $\pm$ 0.11 & 0.38 $\pm$ 0.13 & 0.35 $\pm$ 0.08 & $\mathbf{0.57 \pm 0.13}$ & $\mathbf{0.33 \pm 0.12}$ & $\mathbf{0.70 \pm 0.15}$ & $\mathbf{0.46 \pm 0.14}$& $\mathbf{0.98 \pm 0.01}$& $\mathbf{0.62 \pm 0.37}$\\
\multirow{2}*{} 1.5 & 0.33 $\pm$ 0.11 & 0.49 $\pm$ 0.12 & $\mathbf{0.41 \pm 0.12}$ & $\mathbf{0.59 \pm 0.09}$ & 0.64 $\pm$ 0.13 & 0.71 $\pm$ 0.14 & 0.76 $\pm$ 0.14 & 0.77 $\pm$ 0.12 & 0.98 $\pm$ 0.02 & 0.81 $\pm$ 0.36 \\
\multirow{2}*{} 3.5 & $\mathbf{0.65 \pm 0.12}$ & $\mathbf{0.83 \pm 0.09}$ &$\mathbf{0.55 \pm 0.09}$ & $\mathbf{0.89 \pm 0.04}$ & 0.83 $\pm$ 0.13 & 0.91 $\pm$ 0.06 & 0.87 $\pm$ 0.17 & 0.89 $\pm$ 0.10 & 1.00 $\pm$ 0.00 & 0.91 $\pm$ 0.21 \\
\hline
\end{tabular}
\end{minipage}
\end{table*}

\paragraph{Uncertainty}
The top row of Figure \ref{fig:3} shows the results when uncertainty is present and reputation and intrinsic rewards are not integrated into the system. When compared to results obtained without uncertainty, we note that uncertainty modifies significantly the average cooperation within the cooperative and mixed-motive environments $f \in \{1.5, 3.5\}$, while it does not make a significant difference in the competitive and threshold ones (i.e., $f \in \{0.5, 1.0\}$). The average values and the p-values resulting from the tests are reported in Table \ref{tab:stat_unc_no_unc}. The main effects of uncertainty as follows: 1) in the cooperative setting ($f=3.5$), agents fail to reach cooperative equilibria, causing the average cooperation to drop from $0.98 \pm 0.03$ when uncertainty is absent, to $0.40 \pm 0.07$ in the presence of uncertainty. 2) in mixed-motive games, the average cooperation drops significantly, decreasing from an average value of $0.78 \pm 0.09$ (without uncertainty) to $0.16 \pm 0.06$ (with uncertainty), when $f=1.5$.

\paragraph{Reputation}
The second row of Figure \ref{fig:3} shows the results of the experiments with the inclusion of the reputation mechanism and the effective social norm described in Section \ref{subs:reputationmechanism}. We observe that, except for the competitive environment with $f=0.5$, the final average cooperation presents significant differences compared to the experiments without the reputation mechanism, in particular in the absence of steering agents. We confirm the significance of the results through a t-test comparing the ``Uncertainty'' column in Table \ref{tab:stat_unc_no_unc} with the $0\% - \text{R}$ column in Table \ref{tab:rep_intrinsic}. When steering agents are introduced, the average cooperation of the system increases in all environments: just with $30\%$ presence of steering agents, the average cooperation is significantly higher in the mixed motive and in the cooperative games (see again Tables \ref{tab:stat_unc_no_unc} and \ref{tab:rep_intrinsic}). Notably, the addition of steering agents leads to a slight improvement in cooperation within the competitive scenario as well. This behavior can be explained by remembering that the observed value $f^{obs}_i$ has a lower bound of 0 but no upper bound. Consequently, 
agents encounter a bigger number of observed mixed-motive and cooperative values $f^{obs}_i \in (1,+\infty)$ rather than competitive ones $f^{obs}_i \in (0,1)$. 

\balance
\paragraph{Intrinsic Rewards}
In this section, we describe the results of the experiments where the intrinsic reward formulation of Equation \eqref{eq:intrinsic} is applied, which are reported in the third row of Figure \ref{fig:3}. From the plots can be observed that, even without the addition of reputation and social norms, the average cooperation values are significantly higher in all the environments compared to the case where uncertainty is present but intrinsic rewards are not introduced (top row of Figure \ref{fig:3}). The averages values and the p-values resulting from the tests are contained in Table \ref{tab:stat_unc_no_unc}. 

\paragraph{Reputation and Intrinsic Rewards}
The bottom row of Figure~\ref{fig:3} displays the results of the combined presence of intrinsic rewards and reputation with an effective social norm. In Table \ref{tab:rep_intrinsic} we compare these results with those in which the intrinsic reward is not introduced. Interestingly, for the competitive and threshold environments ($f \in \{0.5,1.0\}$), there is a significant difference among the results when at least $50\%$ or more steering agents are included in the system, namely the combination of steering agents and intrinsic rewards lowers the average cooperation, partially recovering the defective behavior. Thus, in these cases, the formulation of intrinsic rewards, when combined with reputation and steering agents, mitigates the optimistic bias introduced by the presence of steering agents. This effect can be observed by comparing Figures \ref{fig:tr5} and \ref{fig:4tr5}. On the other hand, there are significant differences in the mixed-motive environment for $30\%$ steering agents, and in the cooperative one for cases of $0\%$ and $30\%$, namely the inclusion of intrinsic rewards in the presence of the reputation mechanism enhances cooperation in both these environments. 

\section{Conclusion}
In this work we investigated the impact of uncertain incentive alignment on the cooperative behavior of independent reinforcement learning agents. Our aim was to assess the effectiveness of social mechanisms, such as reputation, steering agents and intrinsic rewards in promoting cooperation in environments marked by this type of uncertainty. First, we observed that uncertainty significantly lowers the agents' cooperative behavior in the cooperative and mixed-motives games, while leaving it unchanged in the competitive and threshold ones. Adding a reputation mechanism with an effective social norm significantly boosts cooperation in the cooperative and mixed-motive environments, even without the addition of steering agents, and without significantly impacting behavior in the competitive and threshold ones. Furthermore, steering agents induce optimism and significantly boost cooperation. The intrinsic reward formulation mimics the effect of reputation without steering agents: the two mechanisms, if introduced separately, yield similar effects. And finally, under uncertainty, the synergy between reputation and intrinsic rewards recovers the competitive behavior in the competitive environment, if a percentage of at least $50\%$ of steering agents is employed. Simultaneously, it promotes cooperation in the mixed-motive and cooperative environments, when no steering agents are in use, or only a small fraction of them are deployed. As future work, we aim to investigate the uncertainty, reputation and social norms on EPGG scenarios involving more than two agents. 
We also plan to explore the role of these mechanisms on more complex simulation environments such as Cleanup and Harvest \cite{SSDOpenSource}, characterised by higher-dimensional action spaces. 
Furthermore, we aim to investigate the combined effects of communication \cite{lazaridou2020emergent, orzan2023} on the emergence of social norms and other social mechanisms (e.g., partner selection) on agents learning under uncertain incentive alignment.


\begin{acks}
This research has been supported by the \href{https://hybrid-intelligence-centre.nl}{Hybrid Intelligence Center}, a 10-year program funded by the Dutch Ministry of Education, Culture and Science through the Netherlands
Organisation for Scientific Research (NWO). Roxana R\u{a}dulescu is supported by the Research Foundation – Flanders (FWO), grant number 1286223N.
\end{acks}



\bibliographystyle{ACM-Reference-Format} 
\bibliography{sample}

\newpage
\appendix
\onecolumn
\section*{Emergent Cooperation under Uncertain Incentive Alignment:}
\label{sec:supp}
\section*{Supplementary Material}
\renewcommand{\thesubsection}{\Alph{subsection}}

\vspace{0.5cm}
\subsection{Learning without Uncertainty: Q-learning Results}

\begin{figure*}[h]
     \centering
\begin{minipage}[t]{\textwidth}
\centering
    \begin{subfigure}[b]{0.23\textwidth}
         \centering
         \includegraphics[width=\textwidth]{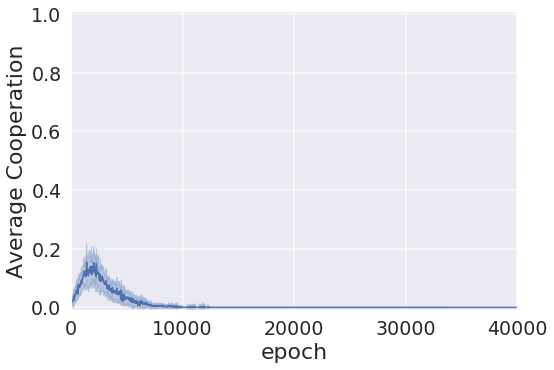}
         \caption{$f=0.5$}
         \label{fig:0tr1}
     \end{subfigure}
    \begin{subfigure}[b]{0.23\textwidth}
         \centering
         \includegraphics[width=\textwidth]{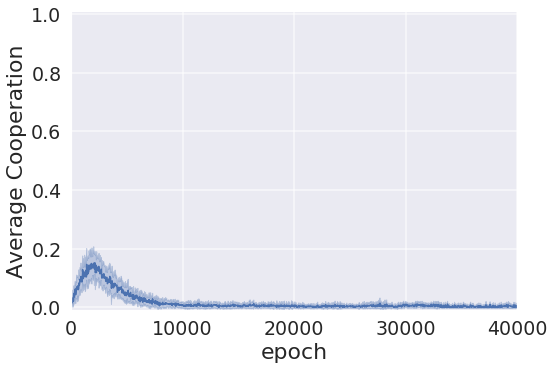}
         \caption{$f=1.0$}
         \label{fig:0tr2}
     \end{subfigure}
      \begin{subfigure}[b]{0.23\textwidth}
         \centering
         \includegraphics[width=\textwidth]{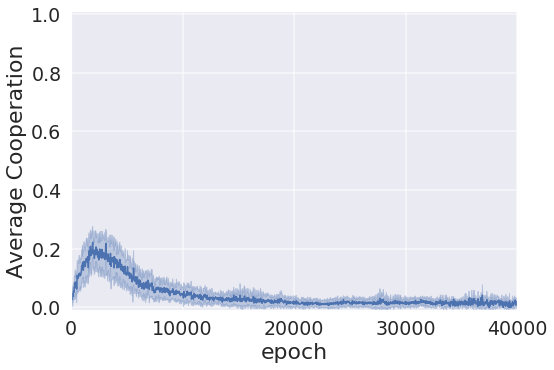}
         \caption{$f=1.5$}
         \label{fig:0tr3}
     \end{subfigure}
       \begin{subfigure}[b]{0.23\textwidth}
         \centering
         \includegraphics[width=\textwidth]{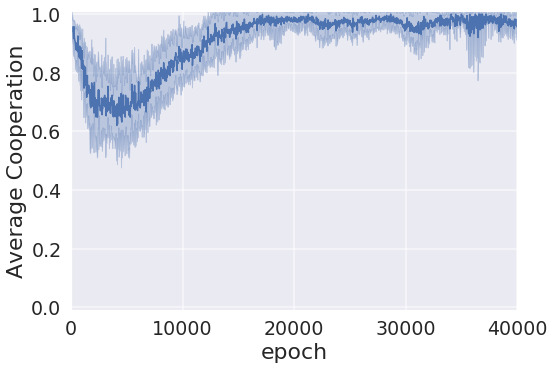}
         \caption{$f=3.5$}
         \label{fig:0tr4}
     \end{subfigure}
\end{minipage}
\begin{minipage}[t]{\textwidth}
\centering
     \begin{subfigure}[b]{0.23\textwidth}
         \centering
         \includegraphics[width=\textwidth]{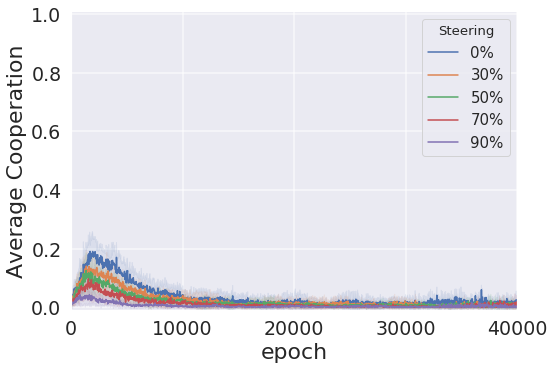}
         \caption{$f=0.5$}
         \label{fig:0tr5}
     \end{subfigure}
     \begin{subfigure}[b]{0.23\textwidth}
         \centering
         \includegraphics[width=\textwidth]{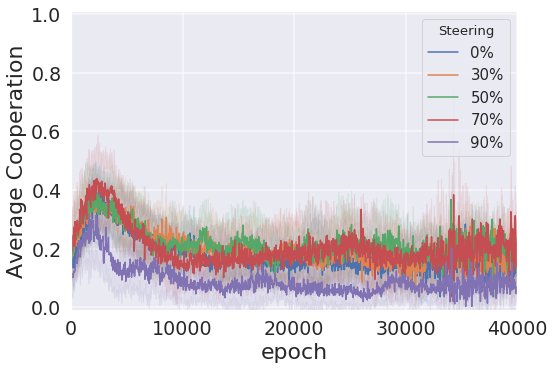}
         \caption{$f=1.0$}
         \label{fig:0tr6}
     \end{subfigure}
    \begin{subfigure}[b]{0.23\textwidth}
         \centering
         \includegraphics[width=\textwidth]{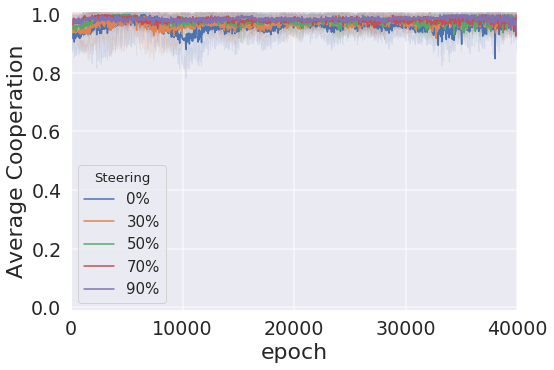}
         \caption{$f=1.5$}
         \label{fig:0tr7}
     \end{subfigure}
     \begin{subfigure}[b]{0.23\textwidth}
         \centering
         \includegraphics[width=\textwidth]{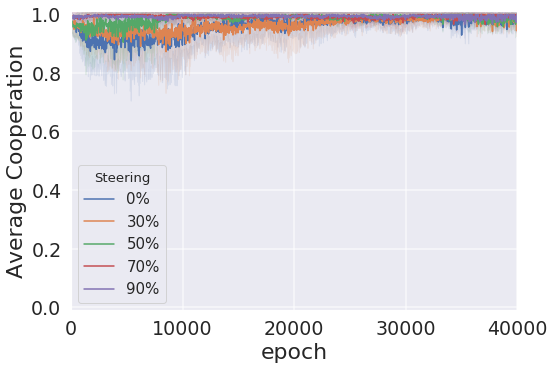}
         \caption{$f=3.5$}
         \label{fig:0tr8}
     \end{subfigure}
\end{minipage}
\begin{minipage}[t]{\textwidth}
\centering
     \begin{subfigure}[b]{0.23\textwidth}
         \centering
         \includegraphics[width=\textwidth]{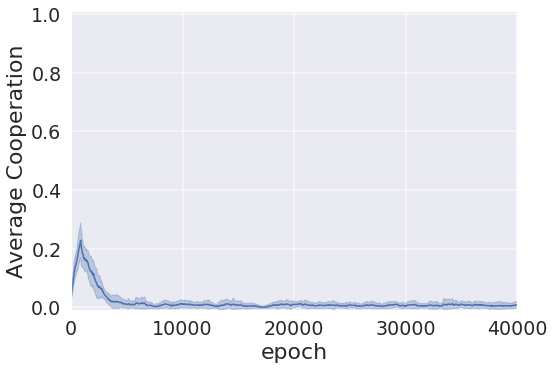}
         \caption{$f=0.5$}
         \label{fig:0tr5}
     \end{subfigure}
     \begin{subfigure}[b]{0.23\textwidth}
         \centering
         \includegraphics[width=\textwidth]{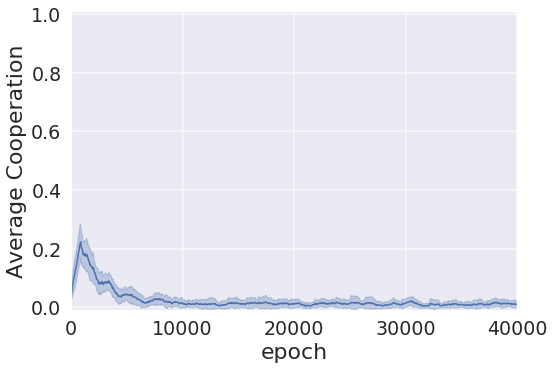}
         \caption{$f=1.0$}
         \label{fig:0tr6}
     \end{subfigure}
    \begin{subfigure}[b]{0.23\textwidth}
         \centering
         \includegraphics[width=\textwidth]{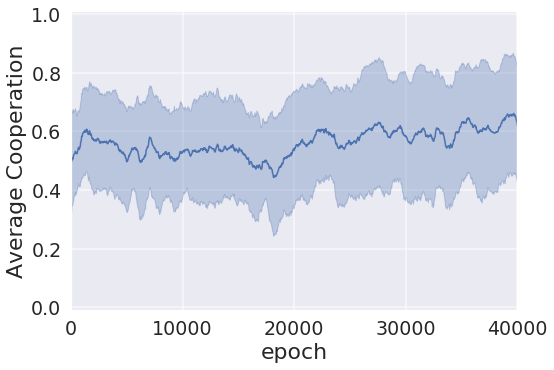}
         \caption{$f=1.5$}
         \label{fig:0tr7}
     \end{subfigure}
     \begin{subfigure}[b]{0.23\textwidth}
         \centering
         \includegraphics[width=\textwidth]{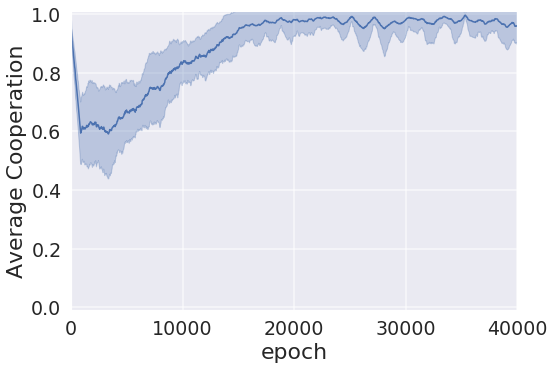}
         \caption{$f=3.5$}
         \label{fig:0tr8}
     \end{subfigure}
\end{minipage}
\begin{minipage}[t]{\textwidth}
\centering
     \begin{subfigure}[b]{0.23\textwidth}
         \centering
         \includegraphics[width=\textwidth]{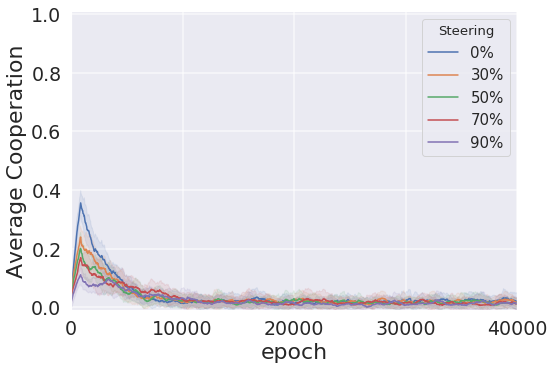}
         \caption{$f=0.5$}
         \label{fig:0tr5}
     \end{subfigure}
     \begin{subfigure}[b]{0.23\textwidth}
         \centering
         \includegraphics[width=\textwidth]{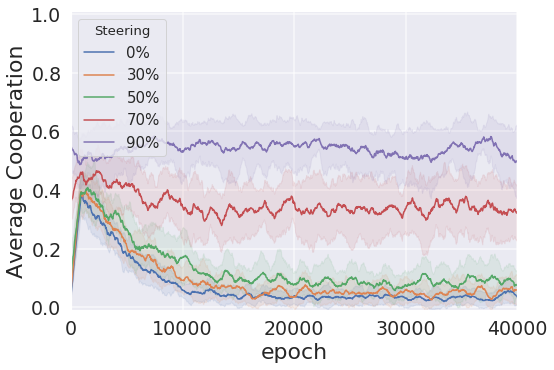}
         \caption{$f=1.0$}
         \label{fig:0tr6}
     \end{subfigure}
    \begin{subfigure}[b]{0.23\textwidth}
         \centering
         \includegraphics[width=\textwidth]{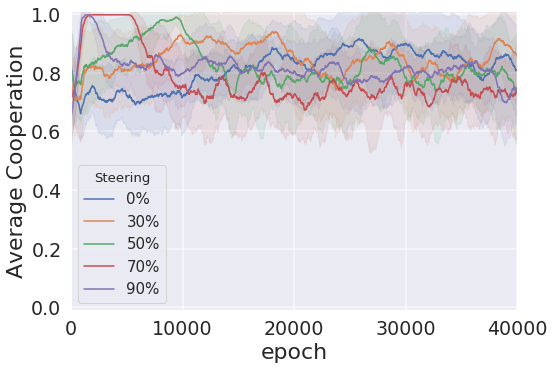}
         \caption{$f=1.5$}
         \label{fig:0tr7}
     \end{subfigure}
     \begin{subfigure}[b]{0.23\textwidth}
         \centering
         \includegraphics[width=\textwidth]{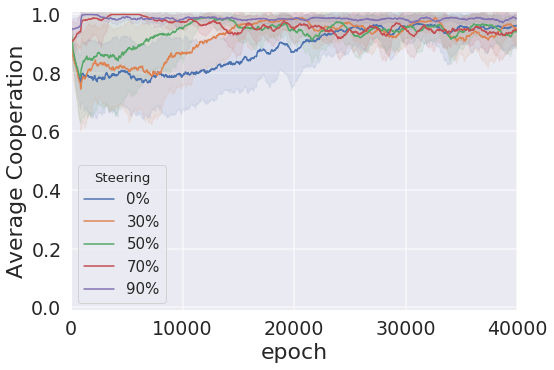}
         \caption{$f=3.5$}
         \label{fig:0tr8}
     \end{subfigure}
\end{minipage}
\caption{Average cooperation for Q-learning agents in four environments with different multiplication factors. Results are presented in four rows: in absence of cooperation-aiding mechanisms (a-d), with the inclusion of the reputation mechanism (e - h), with the intrinsic rewards mechanism (i - l), and with reputation and intrinsic rewards formulation (m - p).}
\label{fig:1}
\end{figure*}

Figure~\ref{fig:1} provides the outcomes of the four experiments outlined in Section \ref{sec:methods}, for tabular Q-learning agents trained on the EPGG in absence of uncertainty. The first row of the figure shows the plots relative the training of Q-learning agents without the addition of reputation mechanism or the intrinsic rewards formulation. We can observe that in the cooperative game ($f=\{3.5\}$) the agents converge to cooperation, while in the competitive and mixed motive games ($f=\{0.5, 1.0, 1.5\}$), they converge to defection. The second row of the plot shows the results in presence of reputation and an effective social norm. In this scenario we can observe that cooperation is quickly reached and stably maintained in the cooperative environment but also in the mixed-motive one, without the need of introducing steering agents. At $f=0.5$ defection remains a stable equilibrium. At $f=1.0$ agents do not perfectly converge to defection, even if the rates of cooperation are very low. The incorporation of steering agents does not significantly change the outcomes. The third row of Figure ~\ref{fig:1} shows the results with the employment of introspective reward, which has an impact only at $m=1.5$: here agents do not converge to defection or cooperation, but reach an average cooperation of $0.51 \pm 0.21$. The fourth row shows the results in presence of the reputation mechanism with a cooperation-aiding social norm, and the intrinsic reward formulation. We let the runs proceed for $40.000$ episodes to let the Q-values stabilize.

\subsection{Learning without Uncertainty: DQN Results}

\begin{figure*}[h]
     \centering
\begin{minipage}[t]{\textwidth}
\centering
    \begin{subfigure}[b]{0.23\textwidth}
         \centering
         \includegraphics[width=\textwidth]{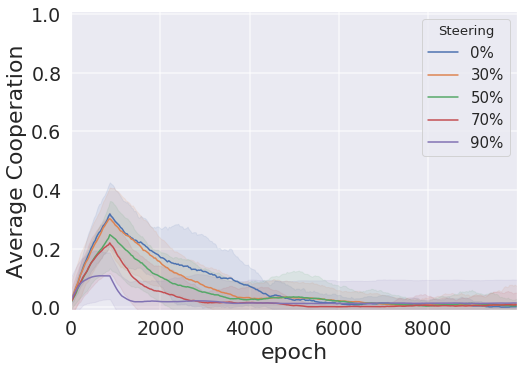}
         \caption{$f=0.5$}
         \label{fig:0tr1}
     \end{subfigure}
    \begin{subfigure}[b]{0.23\textwidth}
         \centering
         \includegraphics[width=\textwidth]{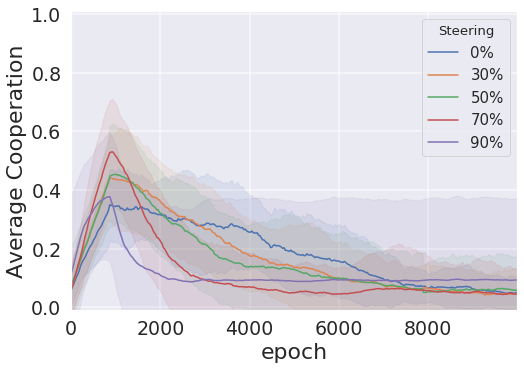}
         \caption{$f=1.0$}
         \label{fig:0tr2}
     \end{subfigure}
      \begin{subfigure}[b]{0.23\textwidth}
         \centering
         \includegraphics[width=\textwidth]{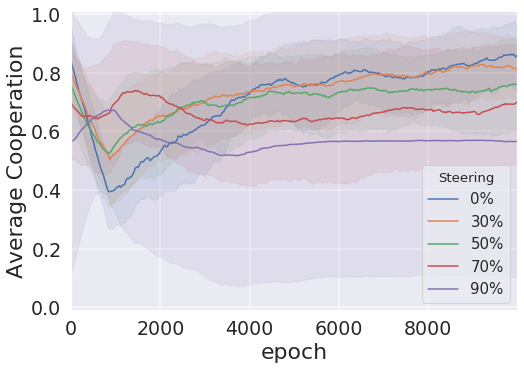}
         \caption{$f=1.5$}
         \label{fig:0tr3}
     \end{subfigure}
       \begin{subfigure}[b]{0.23\textwidth}
         \centering
         \includegraphics[width=\textwidth]{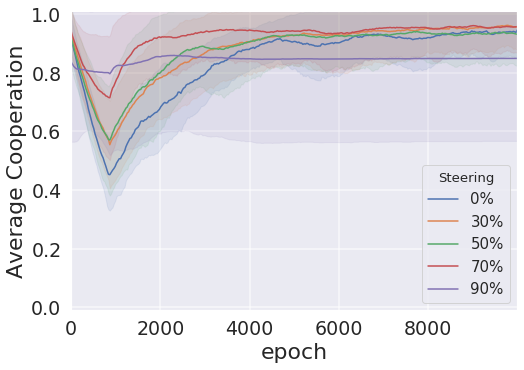}
         \caption{$f=3.5$}
         \label{fig:0tr4}
     \end{subfigure}
\end{minipage}
\caption{Average cooperation for DQN agents trained across environments with different multiplication factors, with the addition of reputation, a social norm that aids cooperation, and the intrinsic rewards formulation.}
\label{fig:2sup}
\end{figure*}

Figure~\ref{fig:2sup} illustrates the results of experiments conducted with DQN agents trained on the EPGG in the absence of uncertainty. In this scenario, we introduce \textbf{reputation} with an effective social norm and apply the \textbf{intrinsic rewards} formulation.  These results show no significant difference compared to the results obtained with intrinsic rewards alone.

\subsection{Learning with Uncertainty: In-depth discussion}
\label{sec:depth}
In this section, we delve into a more extensive discussion of the results obtained from the experiments involving uncertainty, both with and without the three presented social mechanisms (the main results are presented in Section \ref{sec:uncertainty} and illustrated in Figure \ref{fig:3}). In particular, we aim to untangle the reasons underlying the outcomes arising from the interaction among the different cooperation-aiding mechanisms employed.

We observe that when uncertainty is present and intrinsic rewards are employed, the average cooperation values are significantly higher in all the environments compared to the case where uncertainty is still present but intrinsic rewards are disabled (Figure \ref{fig:3}, plots e - h). This happens because agents receive as input the observed multiplication factors instead of the actual one. This uncertainty leads agents to sometimes observe values that hint at a cooperative environment, when instead the game is competitive or mixed-motive. Since we fixed the value of $\beta$ to $0.1$, agents weigh the intrinsic reward more than the real game utility, resulting in higher cooperation values. 

Also employing steering agents alone promotes cooperation in the competitive and threshold environments with uncertainty (Figure \ref{fig:3}, plots i - l). Moreover, the presence of uncertainty brings agents to perceive these environments as cooperative more frequently. Steering agents, given the chosen social norm, push and sustain cooperation every time the perceived multiplication factor is greater or equal to $1$: $f_{obs} > 1$. 

Noteworthy is that the cooperative bias introduced by these two mechanisms occurs as a result of the concurrent learning on environments with varying degrees of incentive alignment. Namely, the presence of cooperative environments impacts positively the agents' behaviour towards cooperation in the threshold and competitive environments.

On the other hand, we observe that combining intrinsic rewards and steering agents alters this dynamic (Figure \ref{fig:3}, plots m - p). In particular, it lowers cooperation in the competitive environment, with respect to the cases in which those mechanisms are used independently.
In this scenario, employing the value $\beta=0.1$, enforces agents to assign more weight to the intrinsic rewards (weighted $0.9$) compared to the rewards coming from the real game (weighted $0.1$), which is played also against steering agents. 
In cooperative and mixed environments, this results in a reduced push towards cooperation from the steering agents. Consequently, because of the effect of concurrent learning, there is a diminished tendency towards cooperation in the threshold and competitive environments as well. Nevertheless, steering agents continue to play a role in enforcing the social norm, pushing learning agents toward competition in the competitive and threshold scenarios.

\end{document}